\documentclass[12pt]{iopart}
\usepackage{graphics,graphicx}
\usepackage{latexsym}
\usepackage{iopams}
\usepackage{color}
\usepackage{pdflscape}

\begin{document}

\title[The Quantemol database]{QDB: a new database of plasma chemistries and
reactions}
\author{Jonathan
  Tennyson$^1$, Sara Rahimi$^2$, Christian Hill$^{1,2}$, Lisa Tse$^2$, Anuradha
Vibhakar$^2$,  Dolica Akello-Egwel$^1$,
Daniel B. Brown$^2$, Anna Dzarasova$^2$, James R. Hamilton$^1$, Dagmar Jaksch$^2$,
Sebastian Mohr$^2$,
Keir Wren-Little$^1$,
Johannes Bruckmeier$^{3}$,
Ankur   Agarwal$^4$,
Klaus   Bartschat$^5$,   
Annemie   Bogaerts$^6$,   
Jean-Paul Booth$^7$,      
Matthew J.  Goeckner$^8$,   
Khaled   Hassouni$^{9}$,    
Yukikazu   Itikawa$^{10}$,       
Bastiaan J Braams$^{11}$,    
E.   Krishnakumar$^{12}$,     
Annarita Laricchiuta$^{13}$, 
Nigel  J.  Mason$^{14}$,        
Sumeet   Pandey$^{15}$,      
Zoran  Lj. Petrovic$^{16}$,       
Yi-Kang   Pu$^{17}$,         
Alok   Ranjan$^{18}$,        
Shahid   Rauf$^{19}$,        
Julian   Schulze$^{20,21}$,     
Miles M.  Turner$^{22}$,       
Peter   Ventzek$^{18}$,   
J. Christopher Whitehead$^{23}$
 Jung-Sik Yoon$^{24}$
}
\address{ 
$^1$          Department of Physics and Astronomy, University College, London, Gower St., London WC1E 6BT, UK\\
$^2$          Quantemol Ltd., University College, London, Gower St., London WC1E 6BT, UK \\
$^{3}$        Infineon Technologies AG \\
$^4$          Applied Materials Inc., 974 E. Arques Avenue, Sunnyvale, CA 94085, USA. \\
$^5$          Department of Physics and Astronomy, Drake University, IA 50311, USA \\
$^6$          Research group PLASMANT, University of Antwerp, Belgium \\
$^7$          Laboratoire de Physique des Plasmas, Ecole Polytechnique, Palaiseau, France\\
$^8$          Department of Physics, University of Texas at Dallas, Richardson TX 78080, USA \\
$^{9}$       Le Laboratoire des Sciences des Proc\'ed\'es et des Mat\'eriaux (LSPM), CNRS-INSIS, France\\
$^{10}$       Institute of Space and Astronautical Science, Sagamihara, Japan \\
$^{11}$       Atomic and Molecular Data Unit, Division of Physical and Chemical Sciences, International Atomic Energy Agency, Vienna, Austria.\\
$^{12}$       Natural Sciences Faculty, Tata Institute of Fundamental Research, Mumbai, India.\\
$^{13}$       PLASMI Lab, CNR NANOTEC Bari , Italy \\
$^{14}$       Department of Physical Sciences, The Open University, UK \\
$^{15}$       Micron Technology Inc., USA\\
$^{16}$       Institute of Physics, University of Belgrade, Serbia\\
$^{17}$       Department of Engineering Physics, Tsinghua University, Beijing, China\\
$^{18}$       TEL Technology Center, America, LLC, USA\\
$^{19}$       Applied Materials Inc., USA. \\
$^{20}$       Institute for Electrical Engineering, Ruhr-University Bochum, Germany\\
$^{21}$       Department of Physics, West Virginia University, USA\\
$^{22}$       National Centre for Plasma Science Technology, Dublin City University, Dublin, Ireland\\
$^{23}$        School of Chemistry, The University of Manchester, UK\\
$^{24}$       Plasma Technology Research Division, National Fusion Research Institute, Gunsan, Korea\\   
 }   
       
       
\ead{$^1$j.tennyson@ucl.ac.uk}
       
\begin{abstract}
  One of the most challenging and recurring problems when modelling plasmas is
  the lack of data on key atomic and molecular reactions that drive
   plasma processes. Even when there are data for some reactions,
  complete and validated datasets of chemistries are rarely available.
  This hinders research on plasma processes and curbs development of
  industrial applications. The QDB project aims to address this
  problem by providing a platform for provision, exchange, and
  validation of chemistry datasets.  A new data model developed for
  QDB is presented. QDB collates published data on both electron
  scattering and heavy-particle reactions.  These data are formed into
  reaction sets, which are then validated against experimental data
  where possible. This process  produces both complete chemistry sets
  and identifies key reactions that are currently unreported in the
  literature. Gaps in the datasets can be filled
  using established theoretical methods. Initial validated chemistry sets for
  SF$_6$/CF$_4$/O$_2$  and SF$_6$/CF$_4$/N$_2$/H$_2$ are presented as examples.
\end{abstract}
 
\submitto{\PSST}
\maketitle

\section{Introduction}

Realistic plasma models of many processes rest on the availability of
reliable atomic and
molecular data, so that the models are able to replicate
the processes that drive
the plasma at the sub\-microscopic level. Particularly for low-temperature
plasmas, which are substantially molecular in composition, the set of
possible processes, which we refer to as reactions below, can be very
large.  For low-temperature plasmas, 
accurate and comprehensive reaction datasets enable complex modelling
of plasma-using technologies that empower our technology-based
society \cite{16BaKu}. Assembling appropriate datasets
is therefore of critical importance.

For a given plasma composition, there are sets of species that are present 
in the plasma and a set of processes, generally called
reactions, that will link the species or different states of the
species. This reaction set is described as the \lq\lq chemistry\rq\rq\
for that plasma. For anything but the simplest molecular plasma, the
number of possible reactions that could make up a chemistry can be
very large \cite{jt440}. The important reactions in a given plasma
will be a subset of all these possible reactions, although it is not
always possible to say in advance precisely which these reactions will
be. In this context it is appropriate to characterise a useful chemistry as
one which has three attributes: 1) The chemistry should be complete, that
is contain all the important reactions for the given plasma. 2) It
should be consistent, that is the reactions should not be unbalanced, thus
resulting in the plasma composition being driven away from the true
composition. 3) Finally, the plasma chemistry should be correct; this
criterion cannot be demonstrated on theoretical grounds alone and
requires validation against experimental measurements made in plasmas.

Assembling plasma chemistries is far from straightforward. While
there may be several chemistries available for relatively simple
systems such as molecular nitrogen plasmas 
\cite{dutuit,13KaLoSt,bultel13,Capitelli2014,16CaCeCo},
they generally do not exist for more complex problems such
as the chemical mixtures typically used in etching and other technological
plasmas. Indeed, given that reactions involving molecular
radicals frequently remain completely uncharacterised \cite{jt646},
it is often a challenge to assemble a complete reaction set for these
chemistries.

Here we present the Quantemol database (QDB).
There are a growing number of databases aimed at supplying the needs
of plasma modellers. For example, the recent LXCat project of
Pitchford {\it et al} \cite{jt647} aims to provide a web-based
platform for data needed to model low-temperature plasmas.
In practice LXCat considers electron collision processes but not
heavy-particle (chemical) reactions.  While both QDB and LXCat are set
up to accept and provide multiple datasets for a single process if
they are available, QDB aims to recommend a dataset for a particular
application while LXCat leaves this choice to its users. 
The Phys4Entry database provides (ro)vibrationally resolved collisional
data, including heterogeneous processes, for modeling re-entry plasmas
\cite{jt628}. For low-temperature,
astronomical plasmas KIDA \cite{jt525,jt604} and BASECOL \cite{jt547}
provide data on chemical reactions and collisional excitation,
respectively.

QDB aims to provide a repository for cross sections and/or rates for
key reactions needed for models of low-temperature, i.e. molecular,
plasmas. QDB collects data on both electron scattering and heavy-particle 
reactions and aims to facilitate and encourage peer-to-peer
data sharing by its users.  At present the data provided are largely
for two-body reactions and hence are appropriate for low-pressure plasmas,
but this will change in the future. Given sets of reactions, QDB then
assembles these sets in chemistries for important plasma mixtures.  If
there are suitable experimental data available, these chemistries can
 be validated. 

The following section gives an overview of QDB, with a technical
specification of the data model given in the appendix. Section 3 catagorises
the process types included in the database while Section 4 summarises
the data sources used. A list of the reactions with a complete set of
references is given as supplementary data to this article. Section 5 
explains our chemistry construction and validation procedure; this is 
illustrated for two chemistries, those comprising
SF$_6$/CF$_4$/O$_2$ and SF$_6$/CF$_4$/N$_2$/H$_2$, respectively. These
chemistries were selected due to their importance in silicon etching.
Section 6 discusses future developments planned for the database,
and the last section provides a summary and conclusions.

\section{Overview}

QDB provides reaction rates, cross sections and
chemistries. The basic data item is the species, which can be state-specified, 
e.g.\ N($^4$S), or not, e.g.\ N$_2$. At
present QDB considers three generic species:
the electron, the photon, and  $M$, the third body in three-body
reactions, plus 405 other atomic and molecular
species. This total rises to 904 species when state-specified
species are counted separately.  

Species can
undergo a series of processes called reactions.  This includes, for
example, elastic scattering or momentum transfer that are not
generally regarded as chemical reactions. These processes are considered in
more detail below. At present the database contains data on 4096
distinct reactions, comprising 2861 energy-dependent cross sections and
2259 temperature-dependent rate coefficients in Arrhenius form.  Note that QDB
allows multiple datasets for the same reaction, and for some reactions
we have distinct data which are available as either cross sections or
rates coefficients. 

Many of these reactions are compiled into chemistries.
Currently QDB has 28 chemistries, which are tabulated and discussed
below. These chemistries can be validated, provided 
appropriate experimental data are available. Currently QDB contains 8
 chemistries with some degree of validation; two of these chemistries are considered in detail below.
Notes on the validation procedure are provided in the form of a datasheet
for each validated chemistry.
Chemistries are awarded a star rating that reflects how far they have
been shown to satisfy the criteria of being complete, consistent, and
correct.  QDB is structured as a MySQL relational
database; the data model used is discussed in the Appendix to this article.

\begin{figure}[bt!]
\centering
\includegraphics[scale=0.3]{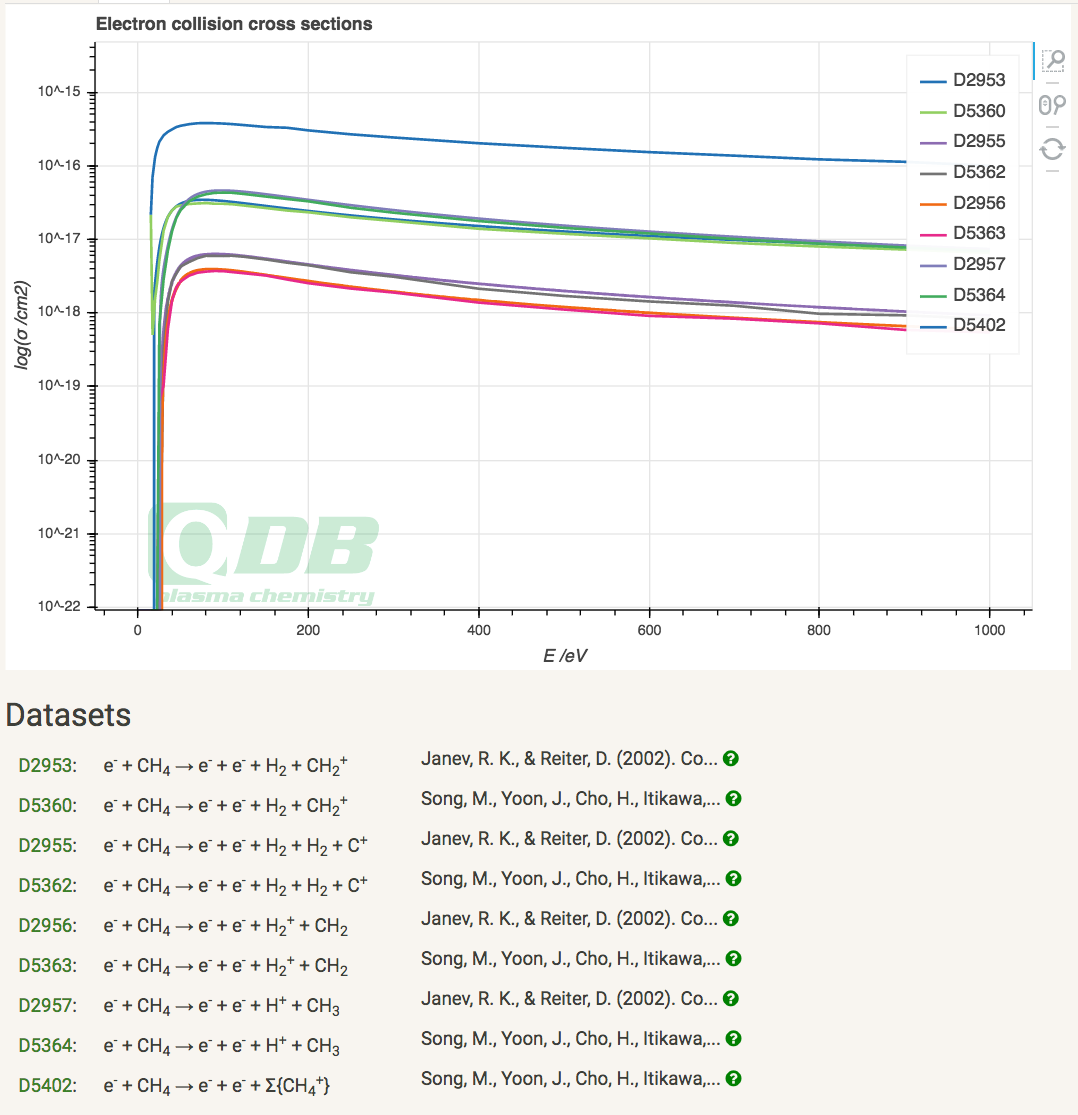}
\caption{Comparison between various electron-impact ionisation 
(EIN, EDI, ETI, see Table 1)
cross sections for methane:  data from Song {\it et al}
\cite{B-381}, and Janev and Reiter \cite{B-354}.}
\label{fig:com_cs} 
\end{figure}

Users can upload new data using the
interface on the QDB website (www.quantemoldb.com) and download
data using a choice of file formats, which is being expanded.  
Currently supported formats are
comma-separated text for each reaction, provided as a zip file or in qdat
format, which facilitates input for Kushner's Hybrid Plasma
Equipment Model (HPEM) \cite{kushner2009hybrid,97GrKu} code.  The zip
format contains all the cross sections, as individual comma-separated
text files, and the rate coefficients that are needed for the
specific chemistry.  It also includes a manifest file listing all the files
provided in the zip archive, a readme file, and a set of citations
in bibtex format. The online view of each dataset contains data sheets and
a form allowing users to provide feedback.
Figure~\ref{fig:com_cs} shows a sample screen shot from QDB giving a comparison
between various cross sections for electron-impact ionisation of methane.
Note that for some reactions, cross sections for the process are obtained
from two different sources; in these cases the agreement is excellent.

Development of QDB is performed with input from the Advisory Board whose members
are co-authors of this paper.

\section{Process types}
Each reaction dataset in QDB is classified as containing 
 cross sections or rate coefficients. The latter can be generated
from the cross sections.
The rate coefficients are expressed in Arrhenius
form stored in the form of three parameters ($A$, $n$, and $E$), which can
be used to compute the rate coefficients at the desired temperatures.
For electron-impact reactions the
Arrhenius formula is
\begin{equation}
A\left(\frac{T_e}{1\;\mathrm{eV}}\right)^n \exp\left( - \frac{E}{T_e} \right)
\end{equation}
 where $T_e$ is the
electron temperature in eV and E is the activation energy in eV.  For heavy-particle 
reactions the Arrhenius formula employed is
\begin{equation}
A (\frac{T_g}{300})^{n}\exp({\frac{-E}{T_g}}),
\end{equation} 
where $T_g$ is the gas temperature in K and $E$ is the activation energy in K.
In both cases, $A$ is the Arrhenius coefficient whose units depend on the
order of the reactions. First-order reactions such as
photodissociation and photoexcitation are  expressed in s$^{-1}$; second-order
reactions, such as electron-impact reactions or two-body
heavy-particle reactions, are expressed in cm$^3$s$^{-1}$; and three-body reactions use
cm$^6$s$^{-1}$.
Cross sections, e.g.\ for electron-neutral-molecule scattering, are given in 
units of cm$^2$ as a function of electron energy in~eV.

Each reaction is classified according to the process considered.
These processes are listed in Tables~\ref{tab:electron_impact},
\ref{tab:heavy_particle}, and \ref{tab:photo} which
consider electron collision processes, heavy-particle reactions and processes
involving photons, respectively. Note that some heavy-particle processes, 
in particular HAS and HIR,
also involve a third body, generically denoted $M$ in the database. The process
label does not depend on the presence of $M$ which is therefore not
included in the process description.

\begin{table}
 \caption {Classification of electron collision processes considered in QDB.}
\begin{center}
\begin{tabular}{lll}
  \hline
Abbreviation&Types of Reaction& Description\\
  \hline
EDX &	deexcitation &	$e + A^* \rightarrow e + A$\\
EEL &	elastic scattering &	$e + A\rightarrow e + A$\\
EIN &	ionization &	$e + A \rightarrow e + A^+ + e$\\
EMT &	momentum transfer &	\\
ERR &	radiative recombination &	$e + A^+ \rightarrow A + h\nu$\\
EDR &	dissociative recombination &	$e + AB^+ \rightarrow A + B$\\
EDS &	dissociation &	$e + AB \rightarrow e + A + B$\\
EDA &	dissociative attachment &	$e + AB \rightarrow A + B^-$\\
EDE &	dissociative excitation &	$e + AB \rightarrow A^* + B + e$\\
EDI &	dissociative ionization &	$e + AB   \rightarrow A^+ + B + 2e$\\
EEX &	electron-impact electronic excitation &	$e + A  \rightarrow e + A^*$\\
ECX &	change of excitation &	$e + A^* \rightarrow e + A^{**}$\\
ERC &	recombination (general) &	$e + A^{+z}  \rightarrow  A^{+(z-1)}$\\
EDT &	electron attachment &	$e + A + B  \rightarrow A + B^-$\\
EVX &	electron-impact vibrational excitation 	&$e + A \rightarrow e + A[v=*]$\\
ETS &	electron total scattering &	$e + A \rightarrow e + \Sigma {A}$\\
ETI &	electron total ionisation &	$e + A \rightarrow e + e + \Sigma {A^+}$\\
ETA &	electron total attachment &	$e + A \rightarrow \Sigma{A^-}$\\
  \hline
\end{tabular}
\label{tab:electron_impact} 
\end{center}
\end{table}

\begin{table}
 \caption {Classification of heavy-particle processes considered in QDB.}
\begin{center}
\begin{tabular}{lll}
  \hline
Abbreviation&Types of Reaction& Description\\
  \hline
HGN &	associative electron detachment &	$A^- + B \rightarrow AB + e$\\
HCX &	charge transfer &	$A^+ + B \rightarrow A + B^+$\\
HIR &	heavy-particle interchange &	$A + BC \rightarrow AB + C$\\
HAS &	association &	$A + B \rightarrow AB$\\
HIN &	heavy-particle collisional ionization &	$A + B \rightarrow A + B^+ + e$\\
HIA &	heavy-particle association and ionization &	$A + B \rightarrow AB^+ + e$\\
HPI &	Penning ionization &	$A + B^* \rightarrow A^+ + B + e$\\
HNE &	neutralization &	$e + B^- \rightarrow B + 2e$\\
HMM &	ions recombination &	$A^-+ B^+ \rightarrow A + B$\\
HDS &	heavy-particle collisional dissociation &	$AB + C \rightarrow A + B + C$\\
HDX &	heavy-particle collisional deexcitation &	$A + B^* \rightarrow A + B$\\
HDN &	heavy-particle dissociative neutralization &	$AB^- + C^+ \rightarrow A + B + C$\\
HDC &	heavy-particle dissociation and charge transfer &	$AB + C^+ \rightarrow A^+ + B + C$\\
HDI &	heavy-particle dissociation and ionization &	$AB + C^* \rightarrow A^+ + B + C$\\
HEX &	heavy-particle excitation 	& $A + B \rightarrow A + B^*$\\
HED &	heavy-particle electron detachment &	 $A^- + B \rightarrow A + B + e$ \\
  \hline
\end{tabular}
\label{tab:heavy_particle}
\end{center}
\end{table}

\begin{table}
 \caption {Classification of photon processes considered in QDB.}
\begin{center}
\begin{tabular}{lll}
  \hline
Abbreviation&Types of Reaction& description\\
  \hline
PDS &	photodissociation &	$AB + h\nu \rightarrow A + B$\\
PEX &	photoexcitation &	$A + h\nu \rightarrow A^*$\\
PRD &	radiative decay &	$A^* \rightarrow A + h\nu$\\
  \hline
\end{tabular}
\label{tab:photo}
\end{center}
\end{table}

\section{Reactions in QDB}

The scientific literature contains many measurements and calculations
of reaction data that provide potentially useful input to plasma
models. However, the task of extracting these data is far from straightforward.
So far our strategy has been to focus on major data compilations and
data sources. A list of those included so far is given in
Table~\ref{t:datain}. In addition a variety of data was taken from
models performed by Kushner and co-workers
\cite{B-59,B-130,B-161,B-172,B-187,B-190,B-193,B-199,B-311,B-312,B-319,B-320,B-331,B-334,B-340,B-357,B-367,B-371}.
These sources were augmented with individual reactions taken 
directly from the original scientific literature. Where no suitable
data could be found, internal (Quantemol) electron collision cross
sections were generated using the Quantemol-N \cite{jt416}
implementation of the UK Molecular R-matrix code (UKRMol)
\cite{jt518}.  As implied by Table~\ref{t:datain}, only a few of these
cross sections have been published, although some have already been
made available via LXCat \cite{jt647} and the Virtual Atomic and
Molecular Data Centre (VAMDC) \cite{jt630}. The process of adding
new data to QDB is a continuous one. The present results represent
a snapshot of the situation as of November 2016.

\begin{landscape}

\begin{table}
 \caption {Data compilations used as input to QDB.}\label{t:datain}
\begin{center}
\begin{tabular}{ll}
\hline
Lead author& System \\
\hline
Itikawa& N$_2$ \cite{B-380}, 
H$_2$O \cite{B-242}, CH$_4$ \cite{B-381}, CO \cite{B-412}, CO$_2$ \cite{B-383}, H$_2$ \cite{B-390}, 
O$_2$ \cite{B-146,B-389}, O \cite{B-257}.\\
Christophorou& CF$_4$ \cite{B-385,B-386}, 
CHF$_3$ \cite{B-236,B-385}, C$_2$F$_6$ \cite{B-385}, 
C$_3$F$_8$ \cite{B-385}, CCl$_2$F$_2$ \cite{B-414}, Cl$_2$ \cite{B-385},
SF$_6$ \cite{B-76,B-273}, C$_2$F$_6$ \cite{B-385}, BCl$_3$ \cite{B-385},
CF$_3$I \cite{B-261}, C$_4$F$_8$ \cite{B-81}\\
Janev &  H$_2$ \cite{B-178,B-384}, C$_X$H$_Y$ \cite{B-354,B-355,B-356}, 
SiH$_4$ \cite{B-303},\\ 
Phelps& N$_2$ \cite{B-142,B-228}, SF$_6$ \cite{B-379}, O$_2$  \cite{B-255} \\
Bartschat& Ar \cite{B-413}, F \cite{B-376}, B \cite{B-277}\\

Quantemol&  BF$_3$ \cite{jt448}, C$_2$H$_2$, C$_3$, C$_3$H$_4$ \cite{jt433}, C$_3$N, CF$_4$, 
CH, CH$_4$ \cite{jt585}, HCN/HNC \cite{jt399},CONH$_3$, COS, CS \cite{jt471}, CaF, F$_2$O,\\
& H$_2$S, HBr, HCHO,  
HCP, Kr, NH$_3$,O$_3$, PH$_3$, SO$_2$, SiF$_2$, SiH$_4$, SiO \cite{jt460}\\
Tennyson& C$_2$ \cite{jt383},  C$_2$OH$_6$ \cite{jt533}, CF \cite{jt315,jt371}, CF$_2$ \cite{jt287,jt371}, CO \cite{jt140,jt527,jt621}, CO$_2$ \cite{jt655},  H$_2$ \cite{jt101,jt235,jt293}, H$_2$O \cite{jt281}, N$_2$ \cite{jt179,jt579},\\   
&N$_2$O \cite{jt204}, NO$_2$ \cite{jt455}, O$_2$ \cite{jt380,jt386,jt543,jt586}  \\
\hline
\end{tabular}
\end{center}
\end{table}

\begin{table}
 \caption {Data sources for electron collision processes included in QDB classified by reaction type.}
.
\begin{center}
\begin{tabular}{ll}
\hline
Reaction & Data source\\
\hline
Deexcitation (EDX)& \cite{B-7,B-201,B-141,B-334,B-269,B-340,B-215,B-312,B-281,B-282,B-187,B-188,B-254}\\

Elastic Scattering (EEL)& \cite{B-128,B-386,B-257,B-390,B-389,B-383,B-141,B-145,B-273,B-282,B-35,B-300,B-172,B-307,B-311,B-312,B-319,B-205,B-334,B-206,B-80,B-210}\\
& \cite{B-340,B-213,B-215,B-88,B-211,B-380,B-219,B-220,B-225,B-227,B-376,B-230,B-232,B-235,B-236,B-239,B-242,B-244,B-119,B-120,B-252,B-381}\\

Ionization (EIN)& \cite{B-128,B-386,B-258,B-389,B-7,B-383,B-141,B-269,B-145,B-147,B-279,B-300,B-307,B-308,B-311,B-185,B-325,B-201,B-203,B-204,B-77,B-334,B-76,B-340,B-212}\\
& \cite{B-119,B-88,B-89,B-215,B-224,B-96,B-97,B-249,B-230,B-99,B-235,B-232,B-357,B-239,B-242,B-244,B-373,B-246,B-376,B-378,B-380,B-253,B-254}\\

Momentum Transfer (EMT)& \cite{B-220,B-385,B-386,B-228,B-389,B-390,B-201,B-76,B-242,B-376,B-381,B-383}\\

Dissociative Recombination (EDR)& \cite{B-130,B-7,B-10,B-12,B-17,B-153,B-35,B-37,B-38,B-172,B-302,B-181,B-205,B-213,B-218,B-351,B-225,B-364}\\
&\cite{B-367,B-368,B-120}\\

Dissociation (EDS)& \cite{B-128,B-386,B-141,B-16,B-147,B-279,B-300,B-172,B-302,B-308,B-76,B-80,B-86,B-88,B-89,B-218,B-93,B-223,B-228}\\
&\cite{B-230,B-236,B-237,B-239,B-242,B-244,B-119,B-379}\\

Dissociative Attachment (EDA)& \cite{B-128,B-386,B-259,B-389,B-262,B-383,B-390,B-300,B-172,B-307,B-323,B-76,B-208,B-209,B-80,B-230,B-236,B-239,B-368,B-242,B-243,B-244,B-119,B-381}\\

Dissociative Excitation (EDE)& \cite{B-216,B-217,B-172}\\

Dissociative Ionization (EDI)& \cite{B-128,B-256,B-386,B-389,B-390,B-261,B-383,B-387,B-388,B-141,B-147,B-172,B-185,B-76,B-77,B-78,B-207,B-81,B-88,B-89,B-224}\\
& \cite{B-96,B-236,B-237,B-381,B-242,B-254,B-249,B-380,B-253,B-382}\\
\hline
\end{tabular}
\label{tab:edata}
\end{center}
\end{table}

\begin{table}
 \caption{Data sources for chemical reactions included in QDB classified by reaction type.}
\begin{center}
\begin{tabular}{ll}
\hline
Reaction & Data source\\
\hline

Associative Electron Detachment (HGN) & \cite{B-98,B-35,B-59,B-239,B-368,B-283,B-31}\\

Charge Transfer (HCX) & \cite{B-130,B-6,B-7,B-135,B-10,B-275,B-20,B-22,B-23,B-25,B-283,B-28,B-31,B-160,B-306,B-171,B-309,B-59}\\
& \cite{B-187,B-188,B-189,B-190,B-320,B-194,B-323,B-80,B-341,B-348,B-100,B-101,B-357,B-103,B-363,B-364,B-239,B-368}\\

Electron-Impact Electronic Excitation (EEX) & \cite{B-128,B-258,B-389,B-390,B-265,B-266,B-267,B-268,B-10,B-142,B-307,B-311,B-312,B-201,B-334}\\
& \cite{B-80,B-209,B-208,B-215,B-227,B-228,B-230,B-235,B-242,B-379,B-374,B-119,B-376,B-250,B-251,B-380,B-381,B-254}\\

Heavy-Particle Interchange (HIR) & \cite{B-10,B-19,B-20,B-26,B-32,B-37,B-38,B-39,B-57,B-58,B-59,B-60,B-63,B-67}\\
& \cite{B-68,B-71,B-73,B-74,B-80,B-101,B-102,B-108,B-112,B-113,B-115,B-116,B-130,B-132,B-153,B-160,B-168,B-172}\\
& \cite{B-191,B-192,B-193,B-195,B-196,B-198,B-199,B-239,B-240,B-283,B-306,B-309,B-313,B-315,B-316,B-317,B-320,B-321}\\
& \cite{B-326,B-327,B-328,B-330,B-333,B-336,B-337,B-338,B-339,B-348,B-357,B-359,B-364,B-365,B-366,B-368,B-372,B-408}\\

Association (HAS) & \cite{B-130,B-6,B-7,B-19,B-20,B-279,B-153,B-281,B-157,B-31,B-291,B-292,B-294,B-295,B-172,B-309,B-313,B-59,B-315}\\
& \cite{B-62,B-318,B-193,B-66,B-196,B-69,B-328,B-331,B-101,B-357,B-359,B-110,B-111,B-239,B-240,B-115,B-372,B-368,B-371}\\

Heavy-Particle Collisional Ionization (HIN) & \cite{B-4,B-7,B-280,B-281,B-152}\\

Heavy-Particle Association And Ionization (HIA) & \cite{B-361,B-5,B-239}\\

Penning Ionization (HPI) & \cite{B-193,B-361,B-362,B-239,B-283,B-187,B-188,B-31}\\

Neutralization (HNE) & \cite{B-230,B-102,B-59,B-147,B-311,B-30,B-31}\\

Ions Recombination (HMM) & \cite{B-130,B-7,B-19,B-23,B-129,B-153,B-36,B-172,B-187,B-59,B-188,B-193,B-322,B-80,B-98,B-106,B-107,B-239,B-368}\\

Heavy-Particle Collisional Dissociation (HDS) & \cite{B-130,B-153,B-6,B-10,B-19,B-20,B-276,B-22,B-151,B-281,B-25,B-283,B-156,B-285,B-287,B-288}\\
& \cite{B-289,B-30,B-313,B-187,B-188,B-190,B-69,B-200,B-357,B-359,B-239,B-240,B-372}\\

Heavy-Particle Collisional Deexcitation (HDX) & \cite{B-320,B-130,B-6,B-359,B-360,B-357,B-331,B-239,B-19,B-22,B-153,B-154}\\

Heavy-Particle Dissociative Neutralization (HDN) & \cite{B-193,B-98,B-36,B-37,B-72,B-264,B-106,B-172,B-239,B-153,B-59}\\

Heavy-Particle Dissociation \&\ Charge Transfer (HDC) & \cite{B-130,B-20,B-25,B-283,B-158,B-160,B-161,B-170,B-172,B-306,B-187,B-59,B-188,B-189}\\
& \cite{B-199,B-80,B-215,B-101,B-357,B-239,B-368,B-240}\\

Heavy-Particle Dissociation And Ionization (HDI) & \cite{B-197}\\

\hline
\end{tabular}
\label{tab:Hdata}
\end{center}
\end{table}
\end{landscape}

A complete list of all reactions currently given in QDB with appropriate
bibliographic references is provided in the supplementary data. 
Tables \ref{tab:edata} and  \ref{tab:Hdata} summarise the sources
of the data  currently available in QDB by process type.  
At present there are relatively few radiative process in the database;
the only ones involve 
radiative decay (PRD) in atoms  \cite{B-187,B-188,B-284}.

\section{Chemistry construction and validation}

\begin{table}
\caption{List and validation status of QDB Chemistries, November 2016}
\begin{center}
\begin{tabular}{lrlc}
\hline
ID & \# reactions & Mixture & Validated?\\
\hline
C3 & 146 &  N$_2$/H$_2$  & \\
C4 & 63 & Ar/H$_2$  & \\
C5 & 155 &  O$_2$/H$_2$  & \\
C6 & 194 & SF$_6$/O$_2$  & Yes\\
C7 & 208 & CF$_4$/O$_2$  & Yes\\
C8 & 61 & SF$_6$  & Yes\\
C9 & 81 & CF$_4$  & Yes\\
C10 & 409 & CF$_4$/O$_2$/H$_2$/N$_2$  & \\
C11 & 193 & C$_4$F$_8$  & Yes\\
C13 & 49 & SiH$_4$  & \\
C14 & 73 & SiH$_4$/NH$_3$  & \\
C15 & 59 & Ar/O$_2$  & \\
C16 & 412 & Ar/O$_2$/C$_4$F$_8$  & Yes\\
C17 & 207 & SiH$_4$/Ar/O$_2$  & \\
C18 & 26 & Ar/Cu  & \\
C19 & 70 & Cl$_2$/O$_2$/Ar  & \\
C20 & 55 & Ar/BCl$_3$/Cl$_2$  & \\
C21 & 238 & Ar/NH$_3$  & \\
C22 & 187 & CH$_4$/H$_2$  & \\
C23 & 81 & C$_2$H$_2$/H$_2$  & \\
C24 & 286 & CH$_4$/NH$_3$  & \\
C25 & 180 & C$_2$H$_2$/NH$_3$  & \\
C26 & 128 & He/O$_2$  & \\
C27 & 562 & CF$_4$/CHF$_3$/H$_2$/Cl$_2$/O$_2$/HBr  & \\
C28 & 590 & CH$_4$/N$_2$  & \\
C29 & 334 & SF$_6$ / CF$_4$/ O$_2$ & Yes\\
C30 & 20 & Ar/Cu/He & \\
C31 & 104 & Ar/NF$_3$ & \\
C32 & 192 &SF$_6$/CF$_4$/N$_2$/H$_2$&Yes \\ 
\hline
\end{tabular}
\label{tab:chemistry-list}
\end{center}
\end{table}

\begin{table}
\caption{Rating scheme for QDB chemistry sets}
\begin{center}
\begin{tabular}{ll}
\hline
Ranking & Description of comparison conditions\\
\hline
1 & Self-consistent but  behaviour differs from  available measurements.\\
2 & Not yet compared; no suitable measurements found.\\
3 & Comparison with measurements for the same process conditions versus one variable:\\
  &  power, gas flow or pressure\\
4 & Comparison with some measurements for different process conditions\\
  &  (more than one comparison) e.g. validation for different pressure regimes. \\
5 & Chemistry tested using more than a program: \\
 &  E.g. Quantemol-P and Quantemol-VT or  another plasma simulation model.\\
\hline
\end{tabular}
\label{tab:rating}
\end{center}
\end{table}

\begin{figure}[bt!]
\centering
{\includegraphics[trim=0.cm 0.0cm 0cm 0.0cm,clip=true,scale=0.085]{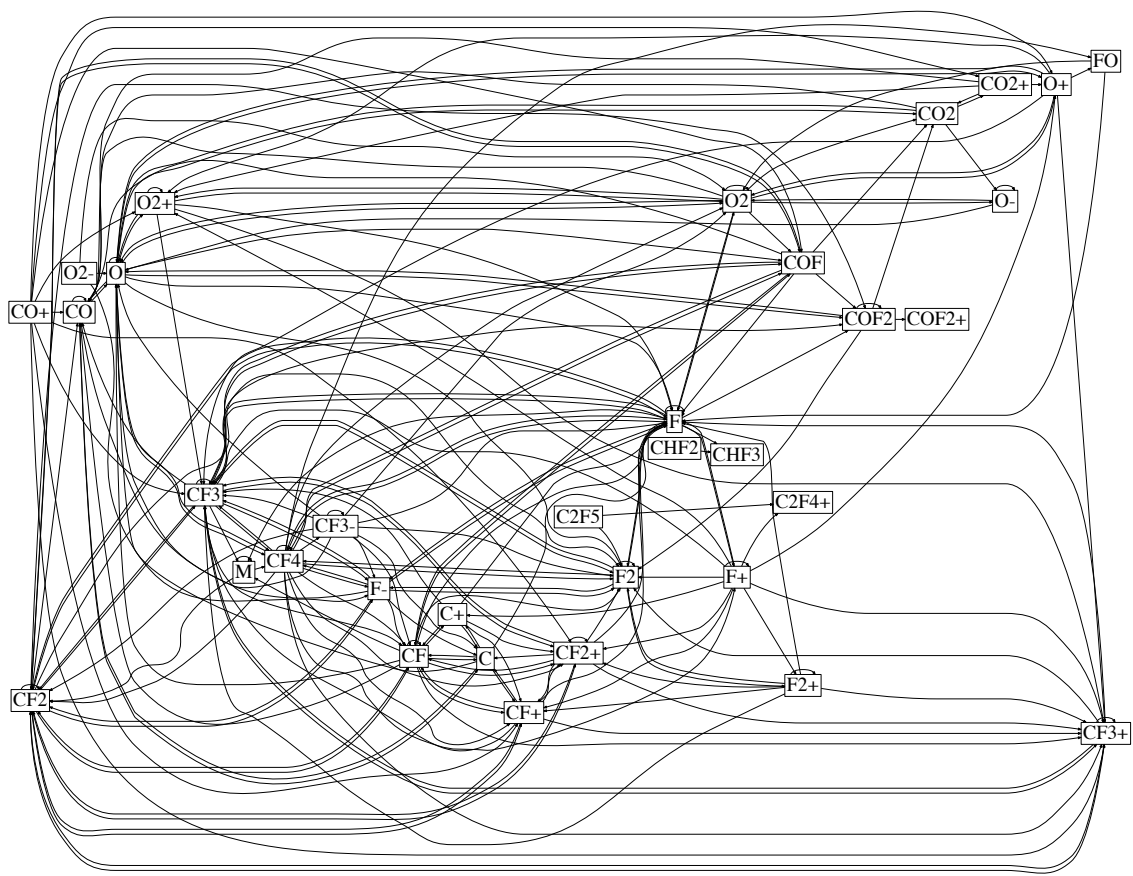}}
\caption{Schematic representation of a chemistry set assembled in QDB for the
CF$_4$/O$_2$ mixture. The species considered are given in the circles, and the lines
give individual reactions linking the various species.}
\label{fig:chem4} 
\end{figure}

The chemistry sets are assembled starting from reactions already
present in QDB; missing reactions are then extracted from the literature and
added to QDB where possible. In cases where important reactions have
not been previously studied, the missing reaction data are calculated
using appropriate methods, such as Quantemol-N \cite{jt416} for electron-molecule
scattering reactions, or by scaling laws, or estimated to provide the
necessary data. These data are added to QDB. This allows us to provide
complete and self-consistent chemistry sets that form the starting
point for validation. Figure~\ref{fig:chem4} illustrates the network
of 196 reactions assembled to characterise the chemistry of SF$_6$ and O$_2$.
Table~\ref{tab:chemistry-list} lists the chemistries avaible in QDB as of
November 2016.

The self-consistency of each chemistry set is checked using a range of
models including Kushner's zero-dimensional GlobalKin
\cite{95GeKu,02DoKu,B-331,B-357} as implemented in Quantemol-P.
GlobalKin couples molecular data (i.e.\ reaction probabilities) with
plasma models to determine plasma properties, such as equilibrium
concentrations of species. A variety of plasmas can be simulated using
this software, such as etching and atmospheric pressure plasma
reactors.
HPEM, as implemented in Quantemol-VT, is also used.

Validation is achieved by using a chemistry set as the basis for the
modelling of different industrial reactors.  Comparison of the model
output with measurement is the principal means by which validation is
achieved. For higher-dimensional simulation, the behaviour of the
species and the surface parameters across the wafer, such as etching or
deposition rates, can also be used for comparison.  Chemistry sets are
given a reliability rating on the basis of these comparisons; see
Table~\ref{tab:rating}.  

We now illustrate this process using two chemistry sets: those for
SF$_6$/CF$_4$/O$_2$ and
SF$_6$/CF$_4$/N$_2$/H$_2$ gas mixture etching Si.  
As discussed below, these sets comprise subsets of mixtures, which
provide important chemistries as well. These were also validated as part
of the validation process. The chemistries were validated
using GlobalKin  and compared with experimental data provided
by Infineon. Unfortunately the data available for complex reactions
sets such as the ones considered is often very limited making the
validation at best partial for these cases. 
We note that the primary goal of these examples is to illustrate the
verification process in a simple form, not to reflect the highest verification
rating. In fact, we consider the examples to only achieve the lowest level of
validation -- that is, agreement in the general trends between simulation and
experiment for one discharge.

The Infineon tool consists of two parts. The first part is a coaxial
microwave plasma discharge.  In this part the electromagnetic wave
propagates along the interface between the plasma column and the
surrounding dielectric tube and the plasma column is sustained by
electromagnetic energy. Free radicals are formed in this chamber with
high efficiency.  This chamber is connected to a larger vessel where
the flux of particles propagates and where the remote wafer to be
etched is located. Our GlobalKin models only attempted to model the
second, larger chamber. GlobalKin performs a spatially homogeneous
plasma chemistry simulation which are coupled with surface reaction
modules. The model uses a Boltzmann solver to obtain electron impact
reaction rate coefficients. 
These models assumed a plasma volume of 90~000 cm$^3$, an area around the plasma of 10~700
cm$^2$. 
The models, which used an assumed diffusion length of 8.3 cm, were initiated using the
feedstock gases. They were run for 500 iterations, corresponding to a total of 1 s, 
which proved sufficient to
reach steady state.

\subsection{SF$_6$/CF$_4$/O$_2$}

\begin{table}[bt]
\caption{Conditions assumed in the model of SF$_6$-CF$_4$-O$_2$ etching.}
 \centering
\begin{tabular}{lc}  
\hline
 Parameter       & Value \\
\hline
 Gas Pressure    & 500 and 700 mTorr \\
 Gas Flow Rate   &                  \\
 SF$_6$             &  800  sccm    \\
 CF$_4$            &   150  sccm    \\
 O$_2$       &   10   sccm     \\
 Power           &   1000 to 2000 W     \\ 
  Substrate       &   Si             \\
\hline    
\end{tabular}
 \label{tab:rep1:parameters}
 \end{table}

 Initially, distinct sets of chemistries for SF$_6$/O$_2$ and CF$_4$/O$_2$
 were constructed and validated separately.  These chemistries were
 then merged and missing reactions, such as SF$_6$ + CF$_3^+
 \rightarrow$ SF$_5^+$ + CF$_4$ ~\cite{babcock1981ion} or SF$_x^-$ +
 CF$_y^+ \rightarrow$ SF$_x$ + CF$_y$, were identified and added.  We
 then set up separate surface chemistries for SF$_6$ and CF$_4$, with
 a focus on silicon etching by F-radicals in the case of SF$_6$ and
 CF$_4$. Surface chemistry parameters were taken from Kokkoris {\it et
   al.} \cite{kokkoris2009global}. Since CF$_4$ formed a smaller
 percentage of the mixture, see Table~\ref{tab:rep1:parameters}, we did
 not include the polymer deposition by CF$_x$ radicals. In addition to
 the F atom reactions, we added reactions of SF$_x$ radicals
 using the same reaction scheme as Kokkoris {\it et al.}
 \cite{kokkoris2009global}, see also comments on this
work by Nelson {\it et al.} \cite{Nelson2012}.  

Only just over 1\%\ of oxygen is added to
 the mixture, as this increases the dissociation of SF$_6$ and CF$_4$.
 Due to its low density, the oxygen-related surface reactions were not
 included in the model.  We also assumed that no significant
 concentration of CS molecules is formed during plasma processing with
 a mixture of SF$_6$ and CF$_4$, due to the large concentration of the
 F radicals in the mixture.  This results in a much higher probability
 of formation of C$_x$F$_y$ and SF$_x$ species.  The conditions of the
 experiments are presented in Table \ref{tab:rep1:parameters}.

To validate the chemistry set, we tested power variation between 1000
and 2000~W at a fixed pressure of 500 mTorr. As a second test we 
varied the  pressure with power fixed at 2000~W.
We used 40\%\ of the
experimental power in our simulation, in order to simulate the energy
dissipation.  This is almost certainly an over estimate of the power
reaching the actual plasma in the Infineon device which explains
why the simulations give faster etch rates than the measurements, see below.

\begin{figure}[bt!]
\centering
{\includegraphics[trim=0.cm 0.0cm 0cm 0.0cm,clip=true,scale=0.4]{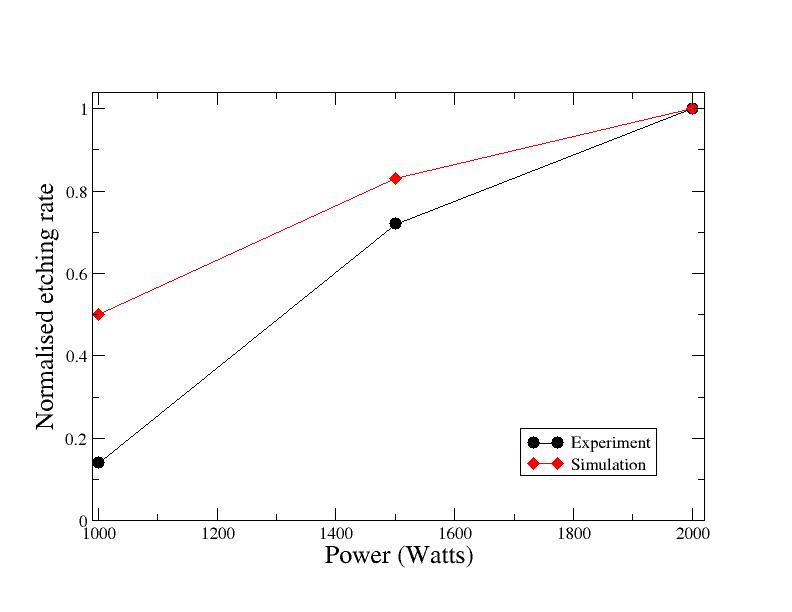}}
{\includegraphics[trim=0.cm 0.0cm 0cm 0.0cm,clip=true,scale=0.4]{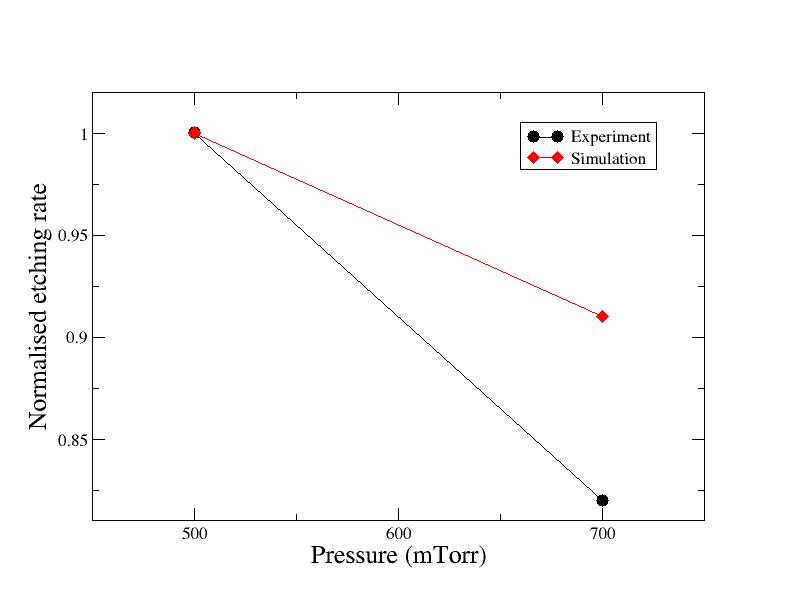}}
\caption{Comparison of the experimental data with the model
results versus power (upper panel) and pressure (lower panel) for Si etching
rate by SF$_6$/CF$_4$/O$_2$.}
\label{fig:O2_power} 
\end{figure}

Our GlobalKin simulations only provide global average
values and therefore cannot provide an absolute quantitative comparison.
For validation of our global model simulation we therefore compare
trends. In particular, 
we compared the trends in the etching rate
with measurements provided by Infineon Technologies, who studied the
effect of both power variation and pressure variation. According to
these measurements, the etch rate increases with increasing power but
decreases with increasing pressure.  Figure~\ref{fig:O2_power}
illustrates the effect of varying power and pressure on the silicon etch rate for a
mixture of SF$_6$/CF$_4$/O$_2$.  Good agreement in the trends is
observed between the results of our simulation and the experimental
data  of Infineon.  By increasing the power in the measurements and
simulation we observe an increase in the Si etching rate but the rate
drops as the pressure is raised.

Given the limited nature of the validation tests we have been able to perform for
this chemistry we give this chemistry set a 3 star rating which is the lowest rating
for a validated chemistry, see Table~\ref{tab:rating}.

\begin{figure}[bt!]
\centering
{\includegraphics[trim=0.cm 0.0cm 0cm 0.0cm,clip=true,scale=0.4]{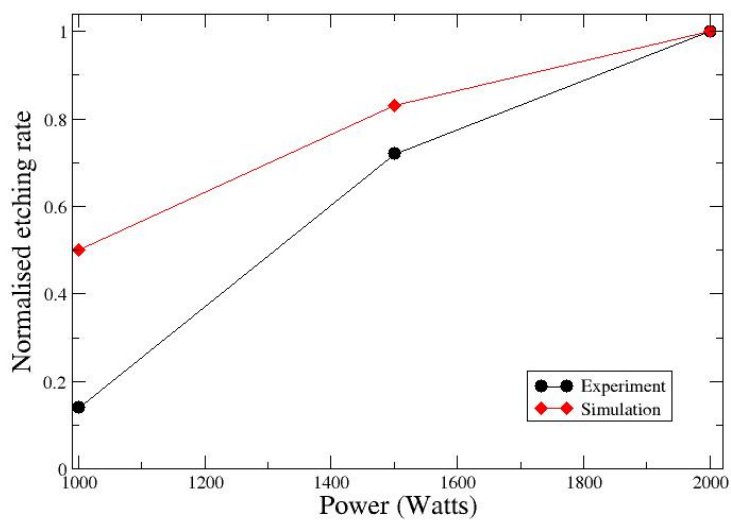}}
\caption{Comparison of the experimental data with the model results versus power for  Si etching
rate by  SF$_6$/CF$_4$/N$_2$/H$_2$.}
\label{fig:N2_power} 
\end{figure}

\subsection{SF$_6$/CF$_4$/N$_2$/H$_2$}
\begin{table*}[bt]
\caption{The conditions for processing in the model for SF$_6$/CF$_4$/N$_2$/H$_2$. }
 \centering
\begin{tabular}{lc}  
\hline
 Parameter       & Value \\ 
\hline
 Gas Pressure    & 500 mTorr\\
 Gas Flow Rate   &                  \\
 SF$_6$            &  800  sccm    \\
 CF$_4$            &   150  sccm    \\
 N$_2$/H$_2$     &   10   sccm     \\
 Power             &    1500 W \\ 
  Substrate        &   Si             \\
\hline
\end{tabular}
 \label{tab:rep2:parameter}
 \end{table*}

 Initially, two sets of chemistries, for SF$_6$/CF$_4$, from the previous validation task,
 and for N$_2$/H$_2$, were constructed and validated separately.
 These chemistries were then merged.  Since there is only a small
 proportion of N$_2$/H$_2$ in the mixture, we excluded species like
 NF$_x$ and CH$_x$F$_y$, as well as the reactions that lead to them.  Rates for ion
 recombination (HMM) reactions such as ${\rm N}^+ + {\rm F}^- \rightarrow
 {\rm N} + {\rm F}$ and ${\rm N}_2^+ {\rm F}^- \rightarrow {\rm N}_2 +
 {\rm F}$ were estimated from Moseley {\it et al} \cite{B-129} and
 added to the chemistry list.  The SF$_6$/CF$_4$ chemistry was
 generated using the same assumptions used  for the SF$_6$/CF$_4$/O$_2$
 mixture.  We also dealt with the surface reactions in a similar fashion to
 SF$_6$/CF$_4$/O$_2$.

\begin{figure}[bt!]
\centering
{\includegraphics[trim=0.cm 0.0cm 0cm 0.0cm,clip=true,scale=0.4]{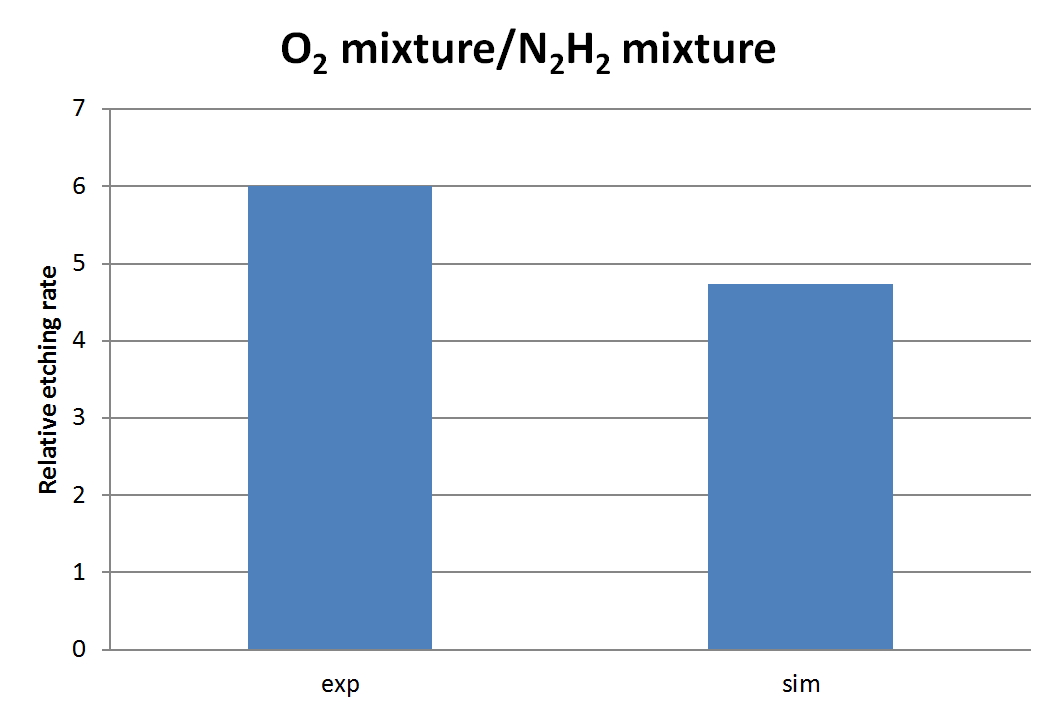}}
\caption{Ratio of etch rate for  Si etching
rate by  mixtures of SF$_6$/CF$_4$/O$_2$ to SF$_6$/CF$_4$/N$_2$/H$_2$:
comparison of measurement with simulation.}
\label{fig:N2H2/O2} 
\end{figure}

Table~\ref{tab:rep2:parameter} summarises the experimental conditions assumed
in the model. 
The experimental data from Infineon we tested against compared etch
 rates, under similar conditions, for the SF$_6$/CF$_4$/N$_2$/H$_2$
 and SF$_6$/CF$_4$/O$_2$ mixtures.  According to the measurements, a
 higher etch rate is found for SF$_6$/CF$_4$/O$_2$ than for
 SF$_6$/CF$_4$/N$_2$/H$_2$. As shown in Fig.~\ref{fig:N2H2/O2}, due
 to the lower dissociation in presence of N$_2$/H$_2$ compared with
 O$_2$, both simulation and experimental results show a higher etch
 rate for the SF$_6$/CF$_4$/O$_2$ mixtures compared to
 SF$_6$/CF$_4$/N$_2$/H$_2$ due to the lower production of F.

Given the limited nature of the validation tests we have been able to perform for
this chemistry we give this chemistry set a 3 star rating which is the lowest rating
for a validated chemistry, see Table~\ref{tab:rating}.

\section{Future developments}

The process of adding both more reactions and chemistries to QDB
is continuous and ongoing.  We will also progressively improve
the validation status of current chemistries and validate more
chemistries, although these activities require appropriate experimental data to
be available for us to validate against. We have developed an initial rating
system for these chemistries, and we plan to develop this further by
allowing users to also submit ratings. At the same time we plan
to implement more formal uncertainty quantification (UQ) procedures
\cite{jt642}.

The processes currently covered by QDB are listed in Tables
\ref{tab:electron_impact}, \ref{tab:heavy_particle}, and
\ref{tab:photo}. These lists do cover all possible low-pressure
gas phase processes. For example, inclusion of vibrationally-resolved reactions for molecules,
such as in electron collisions with CO$_2$ \cite{jt655}, is important for
a number of plasma studies. QDB has the capability to hold such data but
more work on data input and processes considered
will be required to make it fully functional. At present QDB
does not include processes that occur on
surfaces and has only limited data for processes involving a third body. Both of these will be
included in the database in the future. The inclusion of
three-body reactions will extend the coverage to atmospheric pressure
plasmas. 

At present the data can be downloaded in two formats: a generic one and one
that is appropriate for HPEM.  With increasingly large datasets and
sophisticated modelling programs, it is desirable for  data to
be transferred directly from the database to the model using an
application program interface (API).  We plan to develop APIs for
commonly used plasma modelling programs in order to facilitate the use of QDB.
We are also currenly implementing facilities for users to self-assemble
chemistries in their own basket; in the longer term we plan to facilitate
this with an automated chemistry generation tool.


\section{Conclusions}

One of the challenging problems when modelling plasmas is the lack of
reliable chemistry data.  For this purpose, we have developed the
Quantemol Database (QDB),  which aims to provide a platform for
the exchange and validation of reactions that are important in plasmas and
plasma chemistry datasets.  The database
provides data on both electron scattering and heavy-particle reactions,
and it aims to facilitate and encourage peer-to-peer data sharing by its
users.  QDB currently includes almost 5000 reactions and 
28 complete sets of chemistries; so far 8 of these sets have undergone some sort of validation.  
The set of reactions includes more
than 2800 cross sections and more than 2100 sets of reaction rate coefficients in
Arrhenius format for more than 980 species in different states.

\section*{Acknowledgements}
We thank Prof.\ Uwe Czarnetzki, Dr.\ Nobuyuki Kuboi, Prof.\ Satoshi
Hamaguchi, Prof. Mark Kushner and Prof.\ Leanne Pitchford for helpful
discussions during this project. Partial funding was received from the
UK STFC under grants ST/M007774 and ST/K004069, and from the Powerbase
project. Powerbase has received funding from the Electronic Component
Systems for European Leadership Joint Undertaking under grant
agreement No 662133. This Joint Undertaking receives support from the
European Union's Horizon 2020 research and innovation programme and
Austria, Belgium, Germany, Italy, Netherlands, Norway, Slovakia,
Spain, United Kingdom.

\section*{References}

\bibliographystyle{iopart-num}


\providecommand{\newblock}{}

\newpage
\section*{Appendix: The QDB Data Model}

\begin{figure}[h]
\centering
\includegraphics[width=15cm]{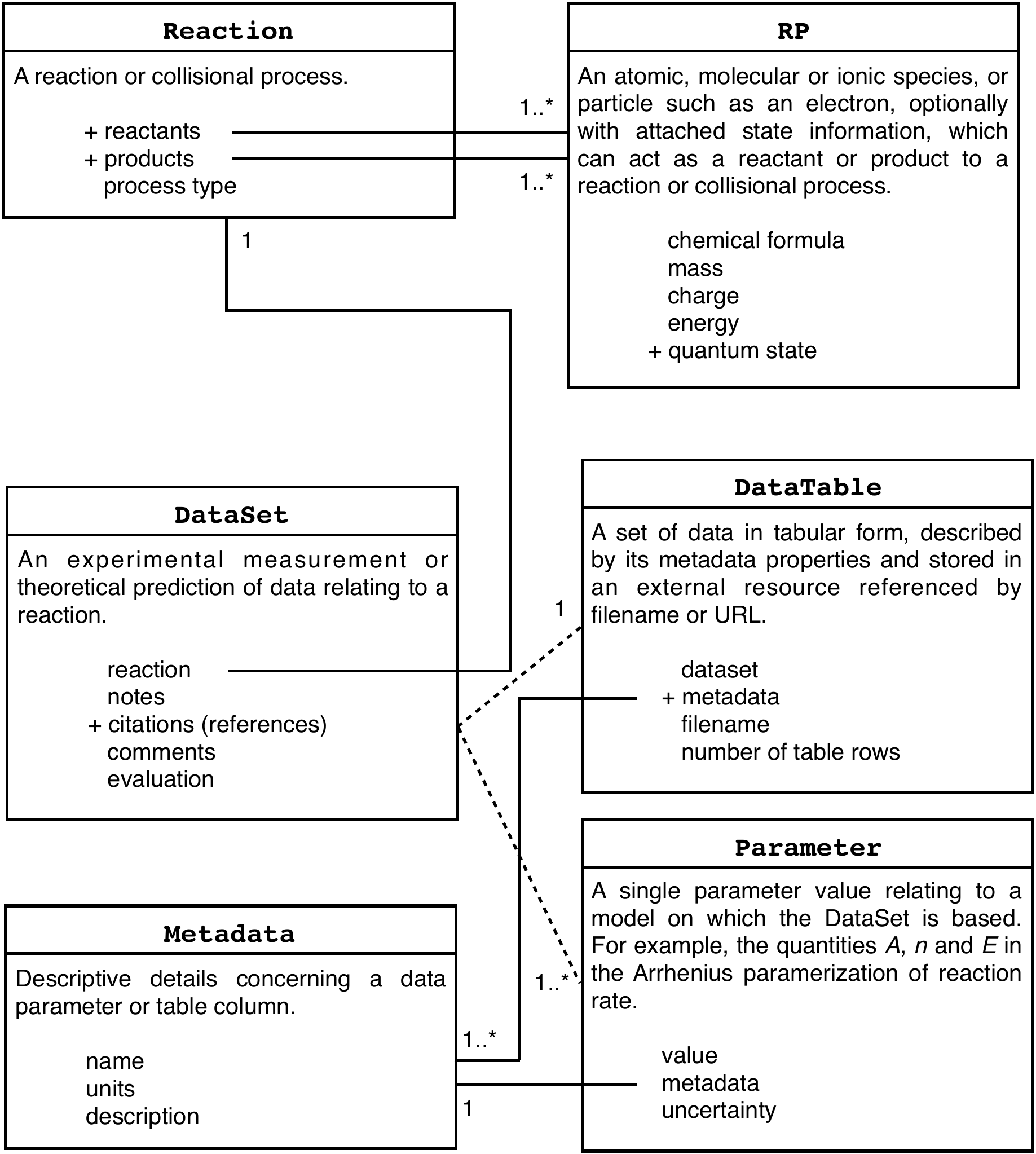}
\caption{An outline of the relations and attributes of the principal tables in the QDB database.}\label{fig:qdb-data-model}
\end{figure}

The Quantemol database (QDB) is implemented using the MySQL relational
database management system (RDBMS). An overview of the principal
tables and their relations is given in Figure
\ref{fig:qdb-data-model}.

Each collision or reaction is considered to take place between one or
more reactants to give one or more products. Each of the reactants and
products may be an atom, ion, molecule, molecular ion, or particle
(such as a photon or an electron), perhaps in a specified quantum
state. These species are represented by their chemical formula
according to a standard notation (for example, \texttt{Ar},
\texttt{H2O}, \texttt{NH2+} for atoms and molecules, $e-$ for electrons
$h\nu$ for photons). State information is attached as a
number of text strings matching a defined pattern, which can be parsed
according to the type of state being considered. A list of some of the
state types with examples is given in Table \ref{tab:qdb-states}.

\begin{table}[h]
\caption{Recognised quantum state designations in QDB} 
\begin{center}
\begin{tabular}{ll}
\hline
Generic excited state & *, **\\
Arbitary key-value pairs & \texttt{n=2}\\
Atomic electronic configuration & \texttt{1s2.2s1}, \texttt{[Ar].3d4}\\
Atomic term symbol & \texttt{3P0}, \texttt{2Po\_1/2}\\
Molecular term symbol & \texttt{X(3}\small{$\Sigma$}\texttt{-g)} \texttt{C(1E"g)}\\
Diatomic vibrational state & \texttt{0}, \texttt{1}\\
Polyatomic vibrational state & \texttt{v1+v2}\\
Racah notation symbol & \texttt{3d[3/2]\_2}\\
Energy, frequency or wavelength & \small{$\lambda$}\texttt{=532 nm}, \texttt{E=12.4 eV}\\
\hline
\end{tabular}
\label{tab:qdb-states} 
\end{center}
\end{table}

Each reactive or collisional process may be described by more than one
DataSet: an experimental measurement or theoretical prediction of rate
data relating to the process. There are two principal types of
DataSet. Cross sections are represented as a table of (electron
energy, cross section value) pairs, stored in an external resource
referenced by filename or URL. Rate data expressed according to an
Arrhenius-like expression, see Eqs.~(1) and (2), are represented by storing separate parameters
$A$, $n$, and $E$. The parameters and the columns of any tabular data
have associated metadata (name, units, and description) in a linked
relational database table.

Different reactive and collisional processes are identified by a
three-letter code (process type) (for example: EDR = dissociative
recombination). The codes employed are an extended version of those
defined in the IAEA document of Humbert {\it et al.} \cite{03HuRaKr}.
A  list of codes is given in Tables 1, 2 and 3; more extensive 
descriptions and examples are given on the
QDB website at http://quantemoldb.com/reactions/processes/.

The structure of the data model allows the user interface to perform
searches of the collisions by species (reactant or product), process
type, and citation. Furthermore, additional fields within the DataSet
table allow for evaluation comments, quality assessment rating, and
validity and usage notes to be stored.

\begin{figure}\label{fig:qdb-chemistry-model}
\begin{center}
\includegraphics[width=7.5cm]{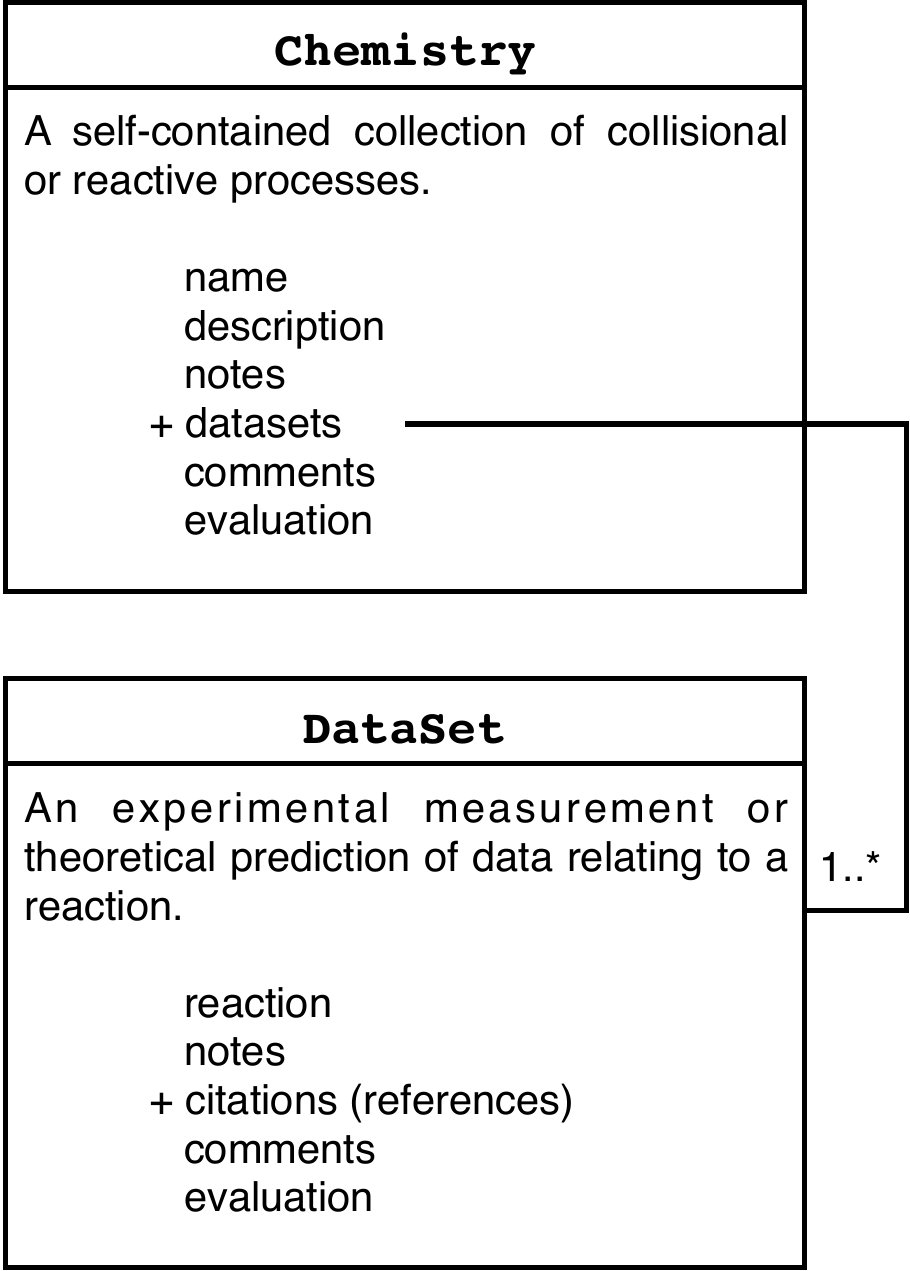}
\caption{An outline of the relations and attributes of the Chemistry and DataSet tables in the QDB database.}
\end{center}
\end{figure}

QDB Chemistries are self-contained collections of collisional and
reactive processes describing the properties of a plasma under some
set of conditions. A table in the relational database holds metadata
relating to each Chemistry, evaluation notes and ratings, and the
associations with the relevant DataSets. See Figure
\ref{fig:qdb-chemistry-model}.

\vspace{2em}

\end{document}